\documentclass[eqsecnum,showpacs,pre,twocolumn]{revtex4}
\usepackage{graphicx}
\usepackage{amsmath}
\usepackage{bm}
\usepackage{epsfig}
\usepackage{amsfonts}
\usepackage{amssymb}
\usepackage[usenames]{color}

\begin{document}

\newcommand{\brm}[1]{\bm{{\rm #1}}}
\newcommand{\Ochange}[1]{{\color{red}{#1}}}
\newcommand{\Ocomment}[1]{{\color{PineGreen}{#1}}}
\newcommand{\Hcomment}[1]{{\color{ProcessBlue}{#1}}}
\newcommand{\Hchange}[1]{{\color{BurntOrange}{#1}}}

\title{{Field theory of directed percolation with long-range spreading}}
\author{Hans-Karl Janssen}
\affiliation{Institut f\"ur Theoretische Physik III, Heinrich-Heine-Universit\"at, 40225
D\"usseldorf, Germany}
\author{Olaf Stenull}
\affiliation{Department of Physics and Astronomy, University of Pennsylvania, Philadelphia
PA 19104, USA}
\date{\today}

\begin{abstract}
\noindent
It is well established that the phase transition between survival and extinction in spreading models with short-range interactions is generically associated with the directed percolation (DP) universality class. In many realistic spreading processes, however, interactions are long ranged and well described by L\'{e}vy-flights, i.e., by a probability distribution that decays in $d$ dimensions with distance $r$ as $r^{-d-\sigma}$. We employ the powerful methods of renormalized field theory to study DP with such long range, L\'{e}vy-flight spreading in some depth. Our results unambiguously corroborate earlier findings that there are four renormalization group fixed points corresponding to, respectively, short-range Gaussian,  L\'{e}vy Gaussian, short-range DP and L\'{e}vy DP, and that there are four lines in the $(\sigma, d)$ plane which separate the stability regions of these fixed points. When the stability line between short-range DP and L\'{e}vy DP is crossed, all critical exponents change continuously. We calculate the exponents describing L\'{e}vy DP to second order in $\varepsilon$-expansion, and we compare our analytical results to the results of existing numerical simulations. Furthermore, we calculate the leading logarithmic corrections for several dynamical observables.
\end{abstract}
\maketitle

\section{Introduction}

The formation and the properties of random structures
have been an exciting topic in statistical physics for many years. In the case
that the formation of such structures obeys local rules, these processes can
often be expressed in the language of epidemic spreading. It is well-known
that two special spreading processes referred to in this language respectively as
simple epidemic with recovery or Gribov process \cite{GrSu78,GrTo79} and
epidemic with removal (general epidemic process) lead to random
structures with the properties of percolation clusters: directed percolation
\cite{BrHa57,CaSu80,Ob80} in the former case and isotropic percolation in the
latter.

The Gribov process, also known in elementary particle physics as Reggeon field
theory (RFT) \cite{Gri67,GM68,Mo78}, is a stochastic multiparticle process
that describes the essential features of a vast number of growth phenomena of
populations without exploitation of the environment near their extinction threshold. The transition between survival and extinction of the population (infected individuals) is a nonequilibrium
continuous phase transition phenomenon and is characterized by universal
scaling laws. The Gribov process belongs to the universality class of local
growth processes with absorbing states \cite{Ja81,Gr82,Ja01} such as the
contact process \cite{Ha74,Li85,JeDi94} and certain cellular automata
\cite{Ki83,Ki85}, and it is relevant to a vast range of models in physics,
chemistry, biology, and sociology. As usual, we refer to this universality class
as the directed percolation (DP)  universality class. For recent reviews see \cite{Hin01,JaTa05}.

A continuum description of DP in terms of a density $n\left(  \mathbf{r}%
,t\right)  $ of infected individuals typically arises from a coarse-graining
procedure in which a large number of microscopic degrees of freedom are averaged
out. Their influence is simply modelled as a Gaussian noise-term in a Langevin
equation. The process has to respect the absorbing state condition: $n\left(
\mathbf{r},t\right)  \equiv0$ is always a stationary state. Then the minimal
stochastic reaction-diffusion equation for the density $n\left(
\mathbf{r},t\right)  $ is constructed as \cite{Ja81}%
\begin{align}
\lambda^{-1}\partial_{t}n\left(  \mathbf{r},t\right)  &=\nabla^{2}n\left(
\mathbf{r},t\right)  -\left[  \tau+\frac{g}{2}\,n\left(  \mathbf{r},t\right)
\right]  \,n\left(  \mathbf{r},t\right)
\nonumber\\
&+\zeta\left(  \mathbf{r},t\right)\, . \label{LangGl}%
\end{align}
The Gaussian noise $\zeta\left(  \mathbf{r},t\right)  $ must also respect the
absorbing state condition, whence
\begin{equation}
\overline{\zeta\left(  \mathbf{r},t\right)  \zeta\left(  \mathbf{r}^{\prime
},t^{\prime}\right)  }=\lambda^{-1}g^{\prime}\,n\left(  \mathbf{r},t\right)
\,\delta\left(  \mathbf{r}-\mathbf{r}^{\prime}\right)  \,\delta\left(
t-t^{\prime}\right)  \label{LangNoise}%
\end{equation}
up to subleading contributions. The history of the process in space and time
defines directed percolation clusters in a ($d+1$)-dimensional space. The
minimal process defined by Eqs.~(\ref{LangGl}) and (\ref{LangNoise}) contains all
the relevant terms needed for a proper field theoretic description of DP.

In realistic situations the infection can be also long-ranged. One may think,
e.g., of an orchard where flying parasites contaminate the trees
practically instantaneous in a widespread manner if the timescale of the
flights of the parasites is much shorter than the mesoscopic timescale of the
epidemic process itself. Thus, following a suggestion of Mollison \cite{Mo77},
Grassberger \cite{Gra86} introduced a variation of the epidemic processes with
an infection probability distribution $P(\mathbf{r})$ which decays with the
distance $r$ as a power law, $P(\mathbf{r})\sim r^{-d-\sigma}$. We will somewhat casually refer to
 such long-range infection as L\'{e}vy-flights although a true
L\'{e}vy-flight is defined via its Fourier transform as $\tilde{P}%
(\mathbf{q})\sim\exp(-bq^{\sigma})$ with $0<\sigma\leq2$ (to ensure positiveness of the distribution).

In Fourier space and in a long-wavelength expansion, the Langevin equation~(\ref{LangGl}) can be generalized to account for L\'{e}vy-flights by a term proportional to
$q^{\sigma}n\left(  q,t\right)  $. In the case of
$2-\sigma\equiv2\alpha>0$, the long-wavelength behavior is naively dominated by
this new term. Grassberger calculated critical exponents in a 1-loop
calculation which were discontinuous in the limit $\alpha
\rightarrow+0$, and therefore the applicability of the results was doubtful.
In a former paper \cite{JOWH99}, we have shown by applying the Wilson momentum
shell renormalization group that only two of the critical exponents are independent in long-range DP, and that the critical exponents change continuously when the transition line between long-ranged and short-ranged spreading (with an $\alpha=\alpha_{c}<0$) is crossed. We have also shown that
\begin{equation}
\sigma_{c}=2(1-\alpha_{c})=d+z-\frac{2\beta}{\nu}=z-\eta
\label{alpha_kritisch}%
\end{equation}
exactly, where $\beta$, $\nu$, $\eta$, and $z$ are the usual exponents of short-ranged
DP in $d$ (transversal) dimensions. These results have been confirmed numerically
by Hinrichsen and Howard \cite{HiHo99}. Note that $\eta
_{\lambda}=z-2-\eta$, which corresponds to the Fisher-exponent of equilibrium
critical phenomena, is negativ here \cite{Ja98}.
For a recent review for DP with long-range interactions see~\cite{Hin07}. In this paper we
reconsider the problem using methods of renormalized field theory in
conjunction with an expansion in $\varepsilon$ and $\alpha$.

The remainder of this paper is organized as follows: Section~\ref{sec:model} reviews the field theoretic formulation of DP with L\'{e}vy-flight spreading to set the stage, to provide background information and to establish notation. Section~\ref{sec:model} first reviews the short-range limit of this model and then discusses our field theoretic analysis of the long-range limit. Section~\ref{sec:hybridModel} represents the main part of this paper. It treats in detail the hybrid model for $\alpha=O(\varepsilon)$. Section~\ref{sec:criticalBehavior} builds up on the results of Sec.~\ref{sec:hybridModel} and present results for the critical exponents and logarithmic corrections of various dynamical observables. Section~\ref{sec:concludingRemarks} contains some concluding remarks.

\section{Modelling DP with L\'{e}vy-flight spreading}
\label{sec:model}

To generalize the diffusional infection rate in the Langevin equation~(\ref{LangGl}), we model spreading by writing
\begin{equation}
\left.  \partial_{t}n\left(  \mathbf{r},t\right)  \right\vert _{\inf}=\int
d^{d}r^{\prime}P(\mathbf{r-r}^{\prime})n\left(  \mathbf{r}^{\prime},t\right) \, .
\label{Levy-inf}%
\end{equation}
As a particular model for the positive distribution $P(\mathbf{r})$ which contains
all relevant properties, we use $P(\mathbf{r})=P_{LR}(\mathbf{r}%
)+P_{SR}(\mathbf{r})$ with a short-range contribution $P_{SR}(\mathbf{r}%
)\propto\exp(-r^{2}/a^{2})$ and a L\'{e}vy-flight part $P_{LR}(\mathbf{r}%
)\propto(r^{2}+a^{2})^{-(d+\sigma)/2}$. $a$ denotes a microscopic length
scale which, for simplicity,  is assumed to be equal in both otherwise independent
distributions. Fourier transformation leads to $\tilde{P}_{SR}(\mathbf{q}%
)\propto\exp\left(  -(aq)^{2}/4\right)  $ and $\tilde{P}_{LR}(\mathbf{q}%
)\propto K_{\sigma/2}(aq),$ where $K_{\sigma/2}$ is the modified
Bessel-function with index $\sigma/2$. Long-wavelength expansion leads to
\begin{align}
\tilde{P}(\mathbf{0})-\tilde{P}(\mathbf{q})&=A(aq)^{2}+\frac{B}{2-\sigma
}\left[  (aq)^{\sigma}-(aq)^{2}\right]
\nonumber\\
&+O(q^{4},q^{2+\sigma}%
)\label{Inf-Spread}%
\end{align}
with positive, non-singular, $\sigma$-dependent constants $A$ and $B$.
$\tilde{P}(\mathbf{q})$ shows the two typical terms: diffusion $\sim(aq)^{2}$
and L\'{e}vy-flights $\sim(aq)^{\sigma}$. It is IR-stable, that is $\tilde
{P}(\mathbf{0})-\tilde{P}(\mathbf{q})$ is positive in both regions $\sigma<2$
and $\sigma>2$ if $q\rightarrow0$. Note the characteristic pole at $\sigma=2$
that leads to a logarithmic contribution $\sim-(aq)^{2}\ln(aq)$ to $\tilde
{P}(\mathbf{q})$ for $\sigma\rightarrow2$. In the following we use for $\tilde{P}(\mathbf{0})-\tilde{P}(\mathbf{q})$ its long-wavelength approximation~(\ref{Inf-Spread}). Whereas
$\tilde{P}(\mathbf{0})-\tilde{P}(\mathbf{q})$ is always a positive real
quantity, its long-wavelength approximation changes the sign at a magnitude $q_g$ of the momentum of order $aq_g =O(1)$ if $\sigma>2$ or if $\sigma<2$ and $A<B/(2-\sigma)$ which leads to a pole in the Greens function of the equation of motion. Of course this pole is a non-physical ghost; it arises when the long-wavelength approximation is used in a momentum regime where it is inapplicable. This happens in particular in dimensional regularization where integrations over internal momenta are extended to infinity. Hence, this method may become inconsistent if both types of $q$-dependencies, $(aq)^{2}$ and $(aq)^{\sigma}$, are used in common. We come
back to this question in Sec.~\ref{sec:landauGhost}. In contrast, Wilsons momentum shell
renormalization procedure avoids this dangerous UV-region $q=O(1/a)$ because
all momenta are restricted to the sphere $q\leq O(\mu)$ with $\mu\ll1/a$. This
was the reason for using Wilson's renormalization group in our former
publication \cite{JOWH99}. Note also, that dimensional regularization leads in the present problem like in other similar problems to the so-called triviality problem at the upper critical dimension, i.e., the dimensionally regularized theory misses logarithmic corrections and thus has to be viewed as an effective theory for low momenta.

The stochastic equation of motion of the DP-process with the L\'{e}vy-flight spreading and
short-range diffusion can be written as
\begin{align}
\lambda^{-1}\partial_{t}n\left(  \mathbf{r},t\right)  &=\left[  \nabla
^{2}-c\left(  -\nabla^{2}\right)  ^{1-\alpha}\right]  n\left(  \mathbf{r}%
,t\right)
\nonumber\\
&-\left[  \tau+\frac{1}{2}g\,n\left(  \mathbf{r},t\right)  \right]
n\left(  \mathbf{r},t\right)  +\zeta\left(  \mathbf{r},t\right)
\, ,\label{LangGl-lr}%
\end{align}
where we set $\sigma=2(1-\alpha)$. Here the L\'{e}vy-term on the right side is
defined in Fourier space as $\left(  -\nabla^{2}\right)  ^{1-\alpha}n\left(
\mathbf{r},t\right)  =\int_{q}q^{2\left(  1-\alpha\right)  }n\left(
\mathbf{q},t\right)  \exp\left(  i\mathbf{q\cdot r}\right)  $. In order to
develop a renormalized field theory, it is useful to recast the Langevin
equation (\ref{LangGl-lr}) as a dynamic response functional
\cite{DeDo76,Ja76}
\begin{align}
\mathcal{J}\left[  \widetilde{s},s\right]  &=\int d^{d}r\,dt\,\lambda\tilde
{s}\bigg\{  \lambda^{-1}\partial_{t}+\left[  \tau-\nabla^{2}+c\left(
-\nabla^{2}\right)  ^{1-\alpha}\right]
\nonumber\\
&+\frac{g}{2}\left(  s-\tilde
{s}\right)  \bigg\}  s\,,\label{J-anfang}%
\end{align}
where $s\left(  \mathbf{r},t\right)  \sim n\left(  \mathbf{r},t\right)  $ is
the rescaled density which ensures that $g^{\prime}=g$ and for which the time
inversion symmetry $s\left(  \mathbf{r},t\right)  \leftrightarrow-\tilde
{s}\left(  \mathbf{r},-t\right)  $ (rapidity reversal in RFT) holds. $\tilde{s}\left(  \mathbf{r},t\right) $ is a response field that describes the response when a local particle source
$h\left(  \mathbf{r},t\right)  \geq0$ is added to the Langevin equation~(\ref{LangGl-lr}). At the level of the dynamic response functional, this source leads to an additional term $\int d^{d}r\,dt\, h\left(  \mathbf{r},t\right) \tilde{s}\left(  \mathbf{r},t\right)$ in Eq.~(\ref{J-anfang}). Having the dynamic response functional, correlation and response functions can be computed as
functional averages (path integrals) of monomials of $s$ and $\tilde{s}$ with
weight $\exp\left\{  -\mathcal{J}\right\} $. Throughout this paper, functional integrals are interpreted in the sense of the so-called prepoint-discretization that sets the step function $\theta\left(  t\right) $
equal to zero for $t=0$~\cite{Ja92}. We stress that the usual short-ranged DP-model is
recovered from the general expression of $\mathcal{J}$ simply by setting
$c=0$, or $c=1$ with $\alpha=0$.

As a first step towards the renormalization group (RG) analysis of this model,
we discuss its canonical scaling behavior. Introducing the usual inverse
length scale $\mu$, we readily find $\tilde{s}\sim s\sim\mu^{d/2}$. For $\alpha>0$, the
long-range L\'{e}vy-term $\sim\left(  -\nabla^{2}\right)  ^{1-\alpha}$
naively dominates the usual diffusion term $\sim\nabla^{2}$. Hence, we may
neglect the latter for $\alpha>0$, and we redefine (by rescaling of some
parameters) $c=1$. This produces an inverse time scale $\lambda\mu^{\sigma}$,
 and $\ \tau\sim\mu^{\sigma}$ for the scaling of the control parameter. Moreover, we
obtain $g^{2}\sim\mu^{\bar{\varepsilon}}$, where $\bar{\varepsilon}%
=2\sigma-d=\varepsilon-4\alpha$ (we will reserve the symbol $\varepsilon$ for
the short-range case, i.e., $\varepsilon=4-d$). The naive dimension of the
coupling constant $g$ allows us to identify the upper critical dimension
$d_{c}(\alpha)=4\left(  1-\alpha\right)  =2\sigma$. This boundary separates
trivial (mean-field or Gaussian) from non-trivial long-range behavior if
$\alpha>0$. Of course, the boundary $\alpha=0$, $d>4$ separates the regions with trivial long-range and trivial short-range DP, and the boundary
$\alpha<0$, $d=4$ separates trivial and non-trivial short-range DP.

\section{Short-range and long-range models}
\label{sec:shortLongModels}

We now turn to perturbation theory. In this section, we will first briefly review the short-range model obtained for $c=0$, which has been discussed previously at many places (see \cite{JaTa05} and the references cited therein). Then we will treat, also briefly, the long-range model obtained for $c\to \infty$.
As usual in dynamical field theory, we focus on those correlation and response functions
\begin{align}
G_{N\,\tilde{N}}=\left\langle\lbrack s]^{N}[\tilde{s}]^{\tilde{N}}\right\rangle
\end{align}
that require renormalization due to the presence of ultraviolet (UV) divergences in Feynman diagrams as well as the corresponding one-particle irreducible (1PI) vertex functions with $\tilde{N}$
($N)$ external $\tilde{s}$- ($s$-) legs, $\Gamma_{\tilde{N}\,N}$.
For background on the methods of renormalization theory, we refer to \cite{ZiJu02}.

\subsection{The short-range model: $c=0$}

We first review ordinary DP which is modeled by $\mathcal{J}$ as given in Eq.~(\ref{J-anfang})
with $c=0$ \cite{Ja81,Ja01}. The upper critical dimension is $d_{c}(0)=4$.
Straightforward dimensional analysis shows that there are three superficially
divergent vertex functions: $\Gamma_{1,1}$, $\Gamma_{1,2}=-\Gamma_{2,1}$,
where the last relation follows from time inversion symmetry. In the following, we use the
superscript $\mathring{}$ denote bare (unrenormalized) couplings, and we use the following renormalization scheme to cure the model of its UV divergences
\begin{align}
{\mathring{s}}  &  =Z^{1/2}s\, ,\qquad\;{\mathring{\tilde{s}}}=Z^{1/2}%
{\tilde{s}}\, ,\qquad\;{\mathring{\lambda}}=Z^{-1}Z_{\lambda}\lambda
\, ,\nonumber\\
{\mathring{\tau}}  &  =Z_{\lambda}^{-1}Z_{\tau}\tau+{\mathring{\tau}}%
_{c}\, ,\qquad\;{\mathring{g}}^{2}=G_{\varepsilon}^{-1}Z^{-1}Z_{\lambda}%
^{-2}Z_{u}u\mu^{\varepsilon}\, , \label{RenSchemSR}%
\end{align}
where $G_{\varepsilon}=\Gamma(1+\varepsilon/2)/(4\pi)^{d/2}$ is a convenient
amplitude, $u$ represents the dimensionless coupling constant, and the control
parameter $\tau$ is zero at the critical point. In dimensional regularization,
the critical bare value of the control parameter, ${\mathring{\tau}}_{c}$, is
of the form%
\begin{equation}
{\mathring{\tau}}_{c}={\mathring{g}}^{4/\varepsilon}S(\varepsilon)\,,
\label{tau-crit}%
\end{equation}
where the Symanzik function $S(\varepsilon)$ has simple IR-poles at each
$\varepsilon=2/k$ with $k=1$, $2$, $\ldots$. Hence, ${\mathring{\tau}}_{c}$ is
not a perturbational quantity and is formally zero in the $\varepsilon
$-expansion. Note, however, that minimal renormalization, i.e., dimensional regularization in conjunction with minimal subtraction, does not imply the $\varepsilon$-expansion~\cite{SchDo89}. The
renormalization factors $Z_{...}$ are functions of $u$ and have in
minimal renormalization the expansions%
\begin{equation}
Z_{...}=1+\sum_{n=1}^{\infty}\frac{Y_{...}^{(n)}(u)}{\varepsilon^{n}}\,,\qquad
Y_{...}^{(n)}(u)=\sum_{l=n}^{\infty}\frac{Y_{...,l}^{(n)}}{l}u^{l}\,,
\label{Z-ExpSR}%
\end{equation}
where the $Z_{...}$ are determined in such a way that the perturbation
expansions of renormalized quantities are free of singularities if
$\varepsilon$ goes to zero. They are given to second order by
\cite{Ja81,Ja01,JaTa05}%
\begin{align}
Z  &  =1+\frac{u}{4\varepsilon}+\left(  \frac{7}{\varepsilon}-3+\frac{9}{2}%
\ln\frac{4}{3}\right)  \frac{u^{2}}{32\varepsilon}+O\left(  u^{3}\right)\, ,
\nonumber\\
Z_{\lambda}  &  =1+\frac{u}{8\varepsilon}+\left(  \frac{13}{\varepsilon}%
-\frac{31}{4}+\frac{35}{2}\ln\frac{4}{3}\right)  \frac{u^{2}}{128\varepsilon
}+O\left(  u^{3}\right) \, ,
\nonumber\\
Z_{\tau}  &  =1+\frac{u}{2\varepsilon}+\left(  \frac{1}{\varepsilon}-\frac
{5}{16}\right)  \frac{u^{2}}{2\varepsilon}+O\left(  u^{3}\right) \, ,
\nonumber\\
Z_{u}  &  =1+\frac{2u}{\varepsilon}+\left(  \frac{7}{\varepsilon}-\frac{7}%
{4}\right)  \frac{u^{2}}{2\varepsilon}+O\left(  u^{3}\right)  \, . \label{Z_SR}%
\end{align}
A renormalization group equation (RGE) for the model can be derived in a routine fashion by exploiting the fact that the unrenormalized response and correlation functions have to be independent of the inverse length scale $\mu$ introduced by renormalization. This reasoning leads straightforwardly to the RGE
\begin{equation}
\left[  \mathcal{D}+\frac{N+\tilde{N}}{2}\gamma\right]  G_{N,\tilde{N}}=0\,,
\label{RGE-SR}%
\end{equation}
with an RGE differential operator $\mathcal{D} = \mu \partial / \partial \mu \vert
_{\mathrm{bare}}$ (the $\vert_{\mathrm{bare}}$ indicates that bare quantities are kept fixed while taking the derivates) given by
\begin{equation}
\mathcal{D}=\mu\frac{\partial}{\partial\mu}+\zeta\lambda\frac{\partial
}{\partial\lambda}+\kappa\tau\frac{\partial}{\partial\tau}+\beta_{u}%
\frac{\partial}{\partial u}\,. \label{RGE-D}%
\end{equation}
The RG functions result from the finite logarithmic derivatives of the
renormalization factors%
\begin{equation}
\gamma_{...}=\left.  \mu\frac{\partial\ln Z_{...}}{\partial\mu}\right\vert
_{\mathrm{bare}}=-u\frac{\partial}{\partial u}Y_{...}^{(1)}(u)=-\sum_{l=1}%
^{n}Y_{...,l}^{(n)}u^{l}\,, \label{Log_Diff}%
\end{equation}
as%
\begin{align}
\beta_{u}  &  =\left.  \mu\frac{\partial u}{\partial\mu}\right\vert
_{\mathrm{bare}}=\left(  -\varepsilon+2\gamma_{\lambda}+\gamma-\gamma
_{u}\right)  u\,,\nonumber\\
\zeta &  =\left.  \mu\frac{\partial\ln\lambda}{\partial\mu}\right\vert
_{\mathrm{bare}}=\gamma-\gamma_{\lambda}\, ,
\nonumber\\
\kappa &=\left.  \mu\frac
{\partial\ln\tau}{\partial\mu}\right\vert _{\mathrm{bare}}=\gamma_{\lambda
}-\gamma_{\tau}\,. \label{Def-RGFu}%
\end{align}
Their perturbation expansions are%
\begin{align}
\gamma(u)  &  =-\frac{u}{4}+\left(  2-3\ln\frac{4}{3}\right)  \frac{3u^{2}%
}{32}+O\left(  u^{3}\right) ,\nonumber\\
\zeta(u)  &  =-\frac{u}{8}+\left(  17-2\ln\frac{4}{3}\right)  \frac{u^{2}%
}{256}+O\left(  u^{3}\right) ,\nonumber\\
\kappa(u)  &  =\frac{3u}{8}-\left(  7+10\ln\frac{4}{3}\right)  \frac{7u^{2}%
}{256}+O\left(  u^{3}\right), \nonumber\\
\beta_{u}(u)  &  =\left[  -\varepsilon+\frac{3u}{2}-\left(  169+106\ln\frac
{4}{3}\right)  \frac{u^{2}}{128}+O\left(  u^{3}\right)  \right]  u\, .
\label{RGE-Fu-SR}%
\end{align}
The asymptotic solution of the RGE, Eq.\ (\ref{RGE-SR}), leads to the stable
fixed point $u=u_{\ast}$ with $u_{\ast}$ given by
\begin{equation}
u_{\ast}=u_{\ast}^{DP}(\varepsilon)=\frac{2\varepsilon}{3}\left[  1+\left(
169+106\ln\frac{4}{3}\right)  \frac{\varepsilon}{288}+O\left(  \varepsilon
^{2}\right)  \right]  , \label{u-DP}%
\end{equation}
as the stable solution of $\beta_{u}(u_{\ast})=0$, and to the scaling form
\begin{align}
G_{N,\tilde{N}}(\{\mathbf{r},t\},\tau)&=l^{(N+\tilde{N})(d+\eta_{SR}%
)/2}
\nonumber\\
&\times G_{N,\tilde{N}}(\{l\mathbf{r},l^{z_{SR}}t\},l^{-1/\nu_{SR}}\tau)
\label{Scal-Green}%
\end{align}
of the response and correlation functions, with the three independent critical exponents%
\begin{equation}
\eta_{SR}=\gamma(u_{\ast})\, ,\quad1/\nu_{SR}=2-\kappa(u_{\ast})\,,\quad
z_{SR}=2+\zeta(u_{\ast})\,. \label{Exp-SR}%
\end{equation}
These are the very well known critical exponent for short-range DP.
To second order in $\varepsilon$-expansion, they are given by
\begin{align}
\eta_{SR}  &  =-\frac{\varepsilon}{6}\left[  1+\left(  \frac{25}{288}%
+\frac{161}{144}\ln\frac{4}{3}\right)  \varepsilon+O(\varepsilon^{2})\right], \nonumber\\
z_{SR}  &  =2-\frac{\varepsilon}{12}\left[  1+\left(  \frac{67}{288}+\frac
{59}{144}\ln\frac{4}{3}\right)  \varepsilon+O(\varepsilon^{2})\right],
\nonumber\\
\nu_{SR}  &  =\frac{1}{2}+\frac{\varepsilon}{16}\,\left[  1+\left(  \frac
{107}{288}-\frac{17}{144}\ln\frac{4}{3}\right)  \varepsilon+O(\varepsilon
^{2})\right]   .
\label{Exp-SR-Eps}%
\end{align}

\subsection{The long-range model: $c\rightarrow\infty$}

As we have shown in \cite{JOWH99} by using Wilson's momentum-shell renormalization
group, and as we discussed in the introduction, the discontinuity of short-range and
long-range critical exponents at $\alpha=0$ is spurious and can be remedied. In
renormalized field theory, the key is to recognize \cite{HoNa89} that there is
a region of small $\alpha=O(\varepsilon)$, where a careful analysis of the RG
flow reveals a smooth connection between the $\alpha<0$, $\alpha
=O(\varepsilon)$, and $\alpha>0$ regions. Here, we analyze the last case,
which belongs to the true long-range region. The case $\alpha=O(\varepsilon),$
being a \textquotedblleft hybrid\textquotedblright\ between short-range and
long-range models, will be deferred to the next subsection.

We recall from our discussion at the end of Sec.~\ref{sec:model}  that the upper
critical dimension for $\alpha > 0$ is $d_{c}(\alpha)=4\left(  1-\alpha\right)  =2\sigma$,
and we define $\bar{\varepsilon}=2\sigma-d$, to be distinguished from
$\varepsilon=4-d$. Considering the response functional $\mathcal{J}$,
Eq.~(\ref{J-anfang}) with $c\neq 0$, we see that the operator $\tilde{s}\nabla^{2}s$ is
superficially irrelevant compared to $\tilde{s}\left(  -\nabla^{2}\right)
^{1-\alpha}s$ and may be dropped formally in the limit $c\rightarrow\infty$.
This limit is feasible after the rescaling $\lambda\rightarrow\lambda/c$,
$\tau\rightarrow\tau c$, and $g\rightarrow gc$. The canonical dimensions of
the fields do not change compared to the short range model. However, the inverse
time-scale changes to $\lambda\mu^{\sigma}$, and the canonical dimensions of the remaining
parameters are $\tau\sim\mu^{\sigma}$ and $g\sim\mu^{\bar{\varepsilon}/2}$. As
above $\Gamma_{1,1}$and $\Gamma_{1,2}=-\Gamma_{2,1}$ are superficially
divergent for $\bar{\varepsilon}\rightarrow0$. Moreover, all divergent
contributions to any vertex function are polynomial in the momenta, so that
the operator $\tilde{s}\left(  -\nabla^{2}\right)  ^{1-\alpha}s$ needs no
counterterm. Hence, we use the renormalization scheme%
\begin{align}
{\mathring{s}} &  =\bar{Z}^{1/2}s\, ,\qquad\;{\mathring{\tilde{s}}}=\bar
{Z}^{1/2}{\tilde{s}}\, ,\qquad\;{\mathring{\lambda}}=\bar{Z}^{-1}%
\lambda\, ,\nonumber\\
{\mathring{\tau}} &  =\bar{Z}_{\tau}\tau+{\mathring{\tau}}_{c}\, ,\qquad
\;{\mathring{g}}^{2}=A_{\bar{\varepsilon}}^{-1}\bar{Z}^{-1}\bar{Z}_{u}%
u\mu^{\bar{\varepsilon}}\, ,\label{RenSchemLR}%
\end{align}
which produces the renormalized response functional
\begin{align}
\mathcal{J}_{\mathrm{LR}}&=\int d^{d}r\,dt\,\lambda\tilde{s}\bigg\{
\lambda^{-1}\bar{Z}\partial_{t}+\left[  \bar{Z}_{\tau}\tau+\left(  -\nabla
^{2}\right)  ^{1-\alpha}\right]
\nonumber\\
&+\bar{Z}_{u}^{1/2}\frac{g}{2}\left(
s-\tilde{s}\right)  \bigg\}  s\,. \label{J-LR}%
\end{align}
Here and in the following we use an overbar to distinguish the renormalization factors of the long-range and hybrid models from those of the short range model. $A_{\bar{\varepsilon}}$ is a suitable amplitude whose precise definition will be given later. Here, in
minimal renormalization, ${\mathring{\tau}}_{c}={\mathring{g}}^{2\sigma
/\bar{\varepsilon}}\bar{S}(\bar{\varepsilon})$ with an appropriate Symanzik
function $\bar{S}(\bar{\varepsilon})$ having simple poles at $\bar
{\varepsilon}=\sigma/k$ with $k=1$, $2$, $\ldots$. Note by comparing
Eq.~(\ref{RenSchemSR}) with Eq.~(\ref{RenSchemLR}) that the renormalization schemes for the short-range and the long-range models are of the same form except for the renormalization factor of the kinetic coefficient, which now is $\bar{Z}_{\lambda}=1$. Here, we have the
expansions%
\begin{equation}
\bar{Z}_{...}=1+\sum_{n=1}^{\infty}\frac{\bar{Y...}^{(n)}(u)}{\bar
{\varepsilon}^{n}}\,,\qquad\bar{Y}_{...}^{(n)}(u)=\sum_{n=1}^{\infty}\frac
{\bar{Y}_{...,l}^{(n)}}{l}u^{l}\,.\label{Z-ExpLR}%
\end{equation}
Of course, the functions $\bar{Y}_{...}^{(n)}(u)$ are different from the
functions $Y_{...}^{(n)}(u)$ and have be determined by perturbation theory. Nevertheless, the RG functions for the long-range case can be transcribed from the short-range case, Eqs.~(\ref{RGE-SR}), (\ref{RGE-D}) and (\ref{Log_Diff}) simply by decorating each
RG-function with an overbar and setting $\bar{\gamma}_{\lambda}=0$, i.e.,
\begin{align}
\bar{\beta} &  =\left.  \mu\frac{\partial u}{\partial\mu}\right\vert
_{\mathrm{bare}}=\left(  -\bar{\varepsilon}+\bar{\gamma}-\bar{\gamma}%
_{u}\right)  u\,,\nonumber\\
\bar{\zeta} &  =\left.  \mu\frac{\partial\ln\lambda}{\partial\mu}\right\vert
_{\mathrm{bare}}=\bar{\gamma}\, ,\quad\bar{\kappa}=\left.  \mu\frac{\partial
\ln\tau}{\partial\mu}\right\vert _{\mathrm{bare}}=-\bar{\gamma}_{\tau
}\,.\label{Def-RGFu-LR}%
\end{align}
The scaling form of the response and correlation functions follows from the RGE as
\begin{align}
G_{N,\tilde{N}}(\{\mathbf{r},t\},\tau)&=l^{(N+\tilde{N})(d+\eta_{lR}%
)/2}
\nonumber\\
&\times G_{N,\tilde{N}}(\{l\mathbf{r},l^{z_{LR}}t\},l^{-1/\nu_{LR}}\tau
) \, ,
\label{Scal-Green-LR}%
\end{align}
with the $\eta_{LR}$ and $\nu_{LR}$ given by
\begin{equation}
\eta_{LR}=\bar{\gamma}(u_{\ast})\,,\qquad1/\nu_{LR}=\sigma-\bar{\kappa
}(u_{\ast})\,.\label{Exp-LR}%
\end{equation}
The third exponent, $z_{LR}$, is related to these exponents by the exact relation
\begin{equation}
z_{LR}=\sigma+\eta_{LR}\,.\label{LR-exRel}%
\end{equation}
Hence, we have only two independent critical exponents in the long-range case, viz.\
$\eta_{LR}$ and $\nu_{LR}$.

After this general discussion, we now turn to our actual perturbation calculation. The propagator of the theory reads
$G(\mathbf{q},t)=\theta(t)\exp\left(  -\lambda(\tau+\left\vert \mathbf{q}%
\right\vert ^{\sigma})t\right) $ in wavevector-time representation. Note that the prepoint discretization mandates that we have to use $\theta(0)=0$ throughout~\cite{Ja92}. Note also, that the propagator is free of the ghost-problem, i.e., the exponent is always negative. The 1-loop self-energy with frequency
$\omega$ and wavevector $\mathbf{q}$ as the first contribution to the vertex
function $\Gamma_{1,1}$ reads
\begin{equation}
\Sigma(\mathbf{q},\omega)=\frac{\lambda g^{2}}{2}\int_{\mathbf{p}}\frac
{1}{i\omega/\lambda+2\tau+\left\vert \mathbf{p}+\mathbf{q}/2\right\vert
^{\sigma}+\left\vert \mathbf{p}-\mathbf{q}/2\right\vert ^{\sigma}%
} \, .\label{Selbst-LR}%
\end{equation}
To calculate this self-energy, it is useful to expand in $\mathbf{q}$ and $\omega$ and to use the
identity%
\begin{align}
&2\pi^{-d/2}\Gamma\left(  1+d/2\right)  \int d^{d}p\,f\left(  \left\vert
\mathbf{p}\right\vert ^{\sigma}\right)
\nonumber\\
& =2\pi^{-d/\sigma}\Gamma\left(
1+d/\sigma\right)  \int d^{2d/\sigma}p\,f\left(  \left\vert \mathbf{p}%
\right\vert ^{2}\right)  \,,\label{Ident}%
\end{align}
which leads the calculation of the primitive divergent 1-loop diagrams back to
integrals of the usual known short-range type. We obtain%
\begin{equation}
\Sigma(\mathbf{q},\omega)=-\frac{g^{2}}{4\bar{\varepsilon}}\tau^{-\overline
{\varepsilon}/\sigma}A_{\bar{\varepsilon}}\left\{  i\omega+\frac{2\sigma
}{\left(  \sigma-\bar{\varepsilon}\right)  }\lambda\tau\right\}
+\mathrm{finite}\,, \label{Selbst-LR-sing}%
\end{equation}
where we have displayed only the pole-terms in $\bar{\varepsilon}$. The pole
at $\bar{\varepsilon}=\sigma$ is an IR-pole and can be removed by introducing
a new mass parameter $m\sim\mu$ instead of $\tau$ via introducing $m^{-\sigma
}=\left.  \partial\ln\Gamma_{1,1}(\omega=0,q,\tau)/\partial q^{\sigma
}\right\vert _{q=0}$ (keeping in mind that it is an IR-poles, one may also simply ignore it in the $\bar{\varepsilon}$-expansion). The amplitude $A_{\bar{\varepsilon}}$ is defined by%
\begin{equation}
A_{\bar{\varepsilon}}=\frac{\Gamma\left(  2-\bar{\varepsilon}/\sigma\right)
\Gamma\left(  1+\bar{\varepsilon}/\sigma\right)  }{\Gamma\left(  \sigma
-\bar{\varepsilon}/2\right)  \left(  4\pi\right)  ^{d/2}}\,.\label{Ampl-A}%
\end{equation}
Note, that $A_{\bar{\varepsilon}}$ becomes $G_{\varepsilon}$ if $\sigma
\rightarrow2$. Expanding $(\mu^{\sigma}/\tau)^{\overline{\varepsilon}/\sigma
}/(1-\overline{\varepsilon}/\sigma)$ in $\overline{\varepsilon}$, and using
the renormalization scheme Eq.~(\ref{RenSchemLR}), we arrive at the singular
part of the vertex function $\Gamma_{1\,1}$ in 1-loop approximation%
\begin{equation}
\Gamma_{1,1}(\mathbf{q},\omega)=\left(  \bar{Z}i\omega+\bar{Z}_{\tau}%
\lambda\tau\right)  -\frac{u}{\bar{\varepsilon}}\left(  \frac{i\omega}%
{4}+\frac{\lambda\tau}{2}\right)  +\ldots\,.\label{Vert11-LR-sing}%
\end{equation}
As announced above, there is no singular term proportional to $\left\vert
\mathbf{q}\right\vert ^{\sigma}$. Using the same techniques, we get
\begin{equation}
\Gamma_{1,2}=-\Gamma_{2,1}=\lambda g\left(  \bar{Z}_{u}^{1/2}-\frac{u}%
{\bar{\varepsilon}}\right)\label{Vert12-LR-sing}%
\end{equation}
for the singular part of the other superficially divergent vertex functions.
We read off the 1-loop renormalizations of the long-range model
\begin{align}
\bar{Z}&=1+\frac{u}{4\bar{\varepsilon}}+O(u^{2})\,,\quad\bar{Z}_{\tau}%
=1+\frac{u}{2\bar{\varepsilon}}+O(u^{2})\,,
\nonumber\\
\bar{Z}_{u}&=1+\frac{2u}%
{\bar{\varepsilon}}+O(u^{2})\,.\label{Z_LR}%
\end{align}
The RG functions here have the expansions
\begin{align}
\bar{\zeta}&=\bar{\gamma}=-\frac{1}{4}u+O(u^{2})\,,\quad\bar{\kappa}=\frac
{1}{2}u+O(u^{2})\,,
\nonumber\\
\bar{\beta}&=\left(  -\bar{\varepsilon}+\frac{7}%
{4}u+O(u^{2})\right)  u\, ,\label{RGFu-LR-exp}%
\end{align}
instead of those given in Eq.~(\ref{Def-RGFu-LR}) for the short-range model.
The stable fixed-point is $u_{\ast}=u_{\ast}^{LR}=4\bar{\varepsilon}%
/7+O(\bar{\varepsilon}^{2})$, and the expansions to first order of the long-range critical
exponents are
\begin{equation}
\eta_{LR}=-\frac{\bar{\varepsilon}}{7}+O\left(  \bar{\varepsilon}^{2}\right)
\,,\quad1/\nu_{LR}=\sigma-\frac{2\bar{\varepsilon}}{7}+O\left(  \bar
{\varepsilon}^{2}\right)  \, ,\label{Exp-LR-exp}%
\end{equation}
which should be compared to the short-range exponents given in Eq.\ (\ref{Exp-LR}).
As has to be the case, our 1-loop results~(\ref{Exp-LR-exp}) are in perfect agreement with the 1-loop results for the long-range exponents derived in \cite{JOWH99} by Wilson's method.

\section{The hybrid model: $\alpha=O(\varepsilon)$}
\label{sec:hybridModel}

Here, we turn to the analysis of the key region in $(d,\alpha)$ space,
namely, $\alpha=O(\varepsilon)$. The naive $\alpha\rightarrow0$ limit of the
long-range model presupposes $\varepsilon\ll\alpha$ and hence fails to
resolve the crossover between the SR and LR models which occurs for
$\alpha=O(\varepsilon)$. For both $\alpha$ and $\varepsilon$ small, we follow
the work of Honkonen and Nalimov~\cite{HoNa89}.

\subsection{Renormalization}
\label{sec:hybridModelRen}

Our starting point is the renormalized response functional
\begin{align}
\mathcal{J}\left[  \widetilde{s},s\right]   &  =\int dt\,d^{d}r\,\bar{\lambda
}\tilde{s}\left\{  \bar{Z}\bar{\lambda}^{-1}\partial_{t}+\bar{Z}_{u}%
^{1/2}\frac{\bar{g}}{2}\left(  s-\tilde{s}\right)  \right. \nonumber\\
&  \left.  +\left[  \bar{Z}_{\tau}\bar{\tau}-\bar{Z}_{\lambda}\nabla
^{2}+c\left(  -\nabla^{2}\right)  ^{1-\alpha}\right]  \right\}  s\ ,
\label{J-hybr}%
\end{align}
where we use the renormalization scheme%
\begin{align}
{\mathring{s}}  &  =\bar{Z}^{1/2}s\, ,\qquad{\mathring{\tilde{s}}}=\bar
{Z}^{1/2}{\tilde{s}}\ ,\qquad{\mathring{\lambda}}=\bar{Z}^{-1}\bar{Z}%
_{\lambda}\bar{\lambda}\, ,\nonumber\\
{\mathring{\tau}}  &  =\bar{Z}_{\lambda}^{-1}\bar{Z}_{\tau}\bar{\tau
}+{\mathring{\tau}}_{c}\, ,\quad{\mathring{g}}^{2}=G_{\varepsilon}^{-1}\bar
{Z}_{\lambda}^{-1}\bar{Z}^{-1}\bar{Z}_{u}\bar{u}\mu^{\varepsilon}%
\, ,
\nonumber\\
\mathring{c}&=\bar{Z}_{\lambda}^{-1}w\mu^{2\alpha}\,,
\label{RenSchem-hyb}%
\end{align}
and the abbreviations $\bar{g}=\sqrt{\bar{u}}\mu^{\varepsilon/2}$ and
$c=w\mu^{2\alpha}$. As before, the term $\tilde{s}\left(  -\nabla^{2}\right)
^{1-\alpha}s$ does not need a counter term as long as $\alpha\neq0$.

Using the approach by he Honkonen and Nalimov, we construct our renormalization factors $\bar{Z}_{...}$ by generalizing the minimal renormalzation program that led to  the expansions
(\ref{Z-ExpSR}) and (\ref{Z-ExpLR}). Here, the renormalization factors are now functions of $\bar{u}$ and $w$, and they contain poles of all linear combinations $\delta_{l,k}=l\varepsilon+2k\alpha$ with
$l=1$, $2$, $\ldots$ and $k=0$, $1$, $2$, $\ldots$:%
\begin{equation}
\bar{Z}_{...}=1+\sum_{l=1}^{\infty}\sum_{k=0}^{\infty}\frac{\bar{Y}%
_{...;l,k}^{\left(  1\right)  }}{l\varepsilon+2k\alpha}w^{k}\bar{u}%
^{l}+O\left(  \delta^{-2}\right)  \, , \label{Z-Exp-hybr}%
\end{equation}
where the coefficients $\bar{Y}_{...;l,k}^{\left(  1\right)  }$ can be chosen such that they are
independent of $\varepsilon$ and $\alpha$ if both are of the same order. To explain this, let us consider the primitive divergence of an irreducible
diagram consisting of $V$ vertices $P$ propagators $L$ independent loops, and
$E$ external amputated legs. By definition, a primitive divergence is a divergence that arises if all the $Ld$ inner momentum integrations tend uniformly to infinity. In primitive diagrams, they are the only UV-divergences. In non-primitive diagrams, they are the divergences that remain after all divergences of the renormalization parts of the diagrams
are tamed by lower order counterterms. After time integrations over the
$(V-1)$ time segments each between vertices, any diagram has the
qualitative form%
\begin{align}
I  &  =\int\frac{(d^{d}p)^{L}}{\left(  m^{2}+p^{2}+cp^{\sigma}\right)  ^{V-1}%
}\, \bar{g}^{V}
\nonumber\\
&\sim\sum_{k=0}^{\infty}\binom{1-V}{k}\int\frac{\left(
cp^{\sigma-2}\right)  ^{k}(d^{d}p)^{L}}{\left(  m^{2}+p^{2}\right)  ^{V-1}%
}\, \bar{g}^{V}
\nonumber\\
&  \sim\sum_{k=0}^{\infty}\binom{1-V}{k}\int\frac{(d^{d}p)^{L}}{\left(
m^{2}+p^{2}\right)  ^{V-1+k\alpha}} \, c^{k}\bar{g}^{V}
\nonumber\\
&\sim\sum_{k=0}^{\infty
}C_{L,V,k}\Lambda^{\Delta(L,V,k)}c^{k}\bar{g}^{V}\,. \label{PrimInt}%
\end{align}
Here, the mass $m^{2}$ is a linear combination of the control parameter $\tau$
and frequencies $\omega$ and serves as an IR regulator. $\Lambda$ is a
momentum cutoff, and $\Delta(L,V,k)=dL-2(V-1)-2k\alpha$ is the degree of primitive
divergence of the diagram. Using the topological relations $P=L+(V-1)$ and
$3V=2P+E$, hence $V=(E-2)+2L$, we obtain%
\begin{equation}
\Delta(L,V,k)=2(3-E)-\left(  L\varepsilon+2k\alpha\right)  \,. \label{PrimDiv}%
\end{equation}
The first part, $2(3-E)$, is the superficial divergence of the diagram with
$E$ legs. The second part, $-\left(  L\varepsilon+2k\alpha\right)  $, denotes
the combination which is converted to a simple pole in dimensional
regularization. This pole-term must be eliminated by a counterterm if
both $\varepsilon$ and $\alpha$ become small quantities. Thus, the overall form of the renormalization factor is as given in Eq.~(\ref{Z-Exp-hybr}). Up to now, the constants $\bar{Y}_{...;l,k}^{\left(
1\right)  }$ may still be functions of $\varepsilon$ and $\alpha$. However, if $\alpha
/\varepsilon$ is finite in the limit $\varepsilon\rightarrow0$, we can neglect
this dependencies in the sense of the minimal renormalization. Hence, we can apply this
Honkonen-Nalimov scheme only if $\alpha=O(\varepsilon)$.

Next, we calculate the logarithmic derivatives of the
renormalization factors. Note that in minimal renormalization
the only terms of the $\beta$-functions,  $\beta_{\bar{u}} = \mu\partial\bar{u}/\partial
\mu \vert _{\mathrm{bare}}$ and $\beta
_{w} =  \mu\partial w/\partial\mu \vert _{\mathrm{bare}}$, which contain $\varepsilon$ and $\alpha$ explicitly come from the $\mu$-factors making ${\mathring{g}}^{2}$ and $\mathring{c}$ dimensionless,
(cf.\ Eq.\ (\ref{RenSchem-hyb})), i.e., $\beta_{\bar{u}}=-\varepsilon\bar{u}+\ldots$
and $\beta_{w}=-2\alpha w+\cdots$. Thus, we obtain from
Eq.\ (\ref{Z-Exp-hybr})
\begin{align}
\bar{\gamma}_{...}  &  =\left.  \mu\frac{\partial\ln\bar{Z}_{...}}{\partial\mu
}\right\vert _{\mathrm{bare}}
\nonumber\\
&=-\left(  \varepsilon\bar{u}\frac{\partial
}{\partial\bar{u}}+2\alpha\frac{\partial}{\partial w}\right)  \sum
_{l=1}^{\infty}\sum_{k=0}^{\infty}\frac{\bar{Y}_{...;l,k}^{\left(  1\right)  }%
}{l\varepsilon+2k\alpha}w^{k}\bar{u}^{l}\nonumber\\
&  =-\sum_{l=1}^{\infty}\sum_{k=0}^{\infty}\bar{Y}_{...;l,k}^{\left(  1\right)
}w^{k}\bar{u}^{l}\,, \label{Gamma-hybr}%
\end{align}
which should be compared to Eq.\ (\ref{Log_Diff}) of the short-range case.

Now we can the relate the functions $\bar{\gamma}_{...}(\bar{u},w)$ to the
original DP-functions $\gamma_{...}(u)$, Eq.\ (\ref{Log_Diff}),  pertaining to the
short-range case. To this end, we consider the hybrid response
functional and its renormalizations for $\alpha=0$:%
\begin{align}
\left.  \mathcal{J}\left[  \widetilde{s},s\right]  \right\vert _{\alpha=0}  &
=\int dt\,d^{d}r\,\bar{\lambda}\tilde{s}\bigg\{  \bar{Z}\bar{\lambda}%
^{-1}\partial_{t}+\bar{Z}_{\tau}\bar{\tau}-\left(  \bar{Z}_{\lambda}+w\right)
\nabla^{2}
\nonumber\\
&+\bar{Z}_{u}^{1/2}\frac{\bar{g}}{2}\left(  s-\tilde{s}\right)
\bigg\}  s\, . \label{J-hybr-alpha-0}%
\end{align}
This functional takes on the form of the original short-range DP response
functional,
\begin{align}
\mathcal{J}\left[  \widetilde{s},s\right] &
=\int dt\,d^{d}r\,\lambda\tilde{s}\bigg\{  Z\lambda^{-1}\partial
_{t}+Z_{\tau}\tau-Z_{\lambda}\nabla^{2}
\nonumber\\
&+Z_{u}^{1/2}\frac{g}{2}\left(
s-\tilde{s}\right)  \bigg\}  s\, ,
\end{align}
if we identify the parameters%
\begin{equation}
\lambda=(1+w)\bar{\lambda}\,,\quad\tau=\left(  1+w\right)  ^{-1}\bar{\tau
}\,,\quad u=\left(  1+w\right)  ^{-2}\bar{u} \label{Param-Ident}%
\end{equation}
and the renormalization factors%
\begin{align}
Z\left(  u\right)   &  =\bar{Z}\left(  \bar{u},w\right)  \ ,\quad\left(
1+w\right)  Z_{\lambda}\left(  u\right)  =\bar{Z}_{\lambda}\left(  \bar
{u},w\right)  +w\ ,\nonumber\\
Z_{\tau}\left(  u\right)   &  =\bar{Z}_{\tau}\left(  \bar{u},w\right)
\ ,\quad Z_{u}\left(  u\right)  =\bar{Z}_{u}\left(  \bar{u},w\right)  \ .
\label{Z-Ident}%
\end{align}
The last identifications lead, by comparison of Eq.\ (\ref{Z-Exp-hybr}) with
Eq.\ (\ref{Z-ExpSR}), to the relations%
\begin{equation}
\sum_{k=0}^{\infty}\bar{Y}_{...;l,k}^{\left(  1\right)  }w^{k}=Y_{...,l}%
^{(1)}\left(  1+w\right)  ^{-2l}\,, \label{Vergl1}%
\end{equation}
in the case of $Z$, $Z_{\tau}$, and $Z_{u}$, and to%
\begin{equation}
\sum_{k=0}^{\infty}\bar{Y}_{\lambda;l,k}^{\left(  1\right)  }w^{k}%
=Y_{\lambda,l}^{(1)}\left(  1+w\right)  ^{1-2l}\,, \label{Vergl2}%
\end{equation}
in the case of $Z_{\lambda}$. Collecting, we obtain for the logarithmic derivatives,
Eq.\ (\ref{Gamma-hybr}),
\begin{align}
\bar{\gamma}\left(  \bar{u},w\right)   &  =\gamma\left(  u\right)
\,,
\nonumber\\
\bar{\gamma}_{\tau}\left(  \bar{u},w\right)  &=\gamma_{\tau}\left(
u\right)  \,,\nonumber\\
\bar{\gamma}_{u}\left(  \bar{u},w\right)   &  =\gamma_{u}\left(  u\right)
\,,
\nonumber\\
\bar{\gamma}_{\lambda}\left(  \bar{u},w\right)  &=\left(  1+w\right)
\gamma_{\lambda}\left(  u\right)  \,. \label{Gamma-Ident}%
\end{align}
Using the renormalizations~(\ref{RenSchem-hyb}), the RG functions become,
\begin{align}
\bar{\beta}_{\bar{u}}  &  =\left.  \mu\frac{\partial\bar{u}}{\partial\mu
}\right\vert _{\mathrm{bare}}=\left(  -\bar{\varepsilon}+2\bar{\gamma
}_{\lambda}+\bar{\gamma}-\bar{\gamma}_{u}\right)  \bar{u}\,,
\nonumber\\
\bar{\beta}_{w}&=\left.  \mu\frac{\partial w}{\partial\mu}\right\vert _{\mathrm{bare}%
}=\left(  -2\alpha+\bar{\gamma}_{\lambda}\right)  w\,,
\nonumber\\
\bar{\zeta}  &  =\left.  \mu\frac{\partial\ln\bar{\lambda}}{\partial\mu
}\right\vert _{\mathrm{bare}}=\bar{\gamma}-\bar{\gamma}_{\lambda}\, ,
\nonumber\\
\bar{\kappa}&=\left.  \mu\frac{\partial\ln\bar{\tau}}{\partial\mu}\right\vert
_{\mathrm{bare}}=\bar{\gamma}_{\lambda}-\bar{\gamma}_{\tau}\,.
\label{Def-RGFu-overbar}%
\end{align}
However, Eqs.~(\ref{Param-Ident}) and (\ref{Gamma-Ident}) suggest that it is more appropriate to use instead of  $\bar{\lambda}$, $\bar{\tau}$, $\bar{u}$, and $w$ the parameters $\lambda$, $\tau$, and $u$ defined by Eq.\ (\ref{Param-Ident}), and
\begin{equation}
v=\frac{2\alpha w}{1+w}\,, \label{Def-w}%
\end{equation}
as the parameters of the theory. The response functional then becomes
\begin{align}
\mathcal{J}\left[  \widetilde{s},s\right]  &=\int dt\,d^{d}r\,\lambda\tilde
{s}\bigg\{  Z\lambda^{-1}\partial_{t}+Z_{\tau}\tau-Z_{\lambda}\nabla^{2}%
\nonumber\\
& +\frac{v}{2\alpha}\left[  \mu^{2\alpha}\left(  -\nabla^{2}\right)  ^{1-\alpha
}+\nabla^{2}\right]  +Z_{u}^{1/2}\frac{g}{2}\left(  s-\tilde{s}\right)
\bigg\}  s\ . \label{J-hybr-final}%
\end{align}
Note, that $\mathcal{J}$ coincides up to a rescaling with $\mathcal{J}_{LR}$,
Eq.~(\ref{J-LR}), if $v=2\alpha$. Note also, that the gradient terms
of this response functional take the same form as proposed for the
long-wavelength or gradient expansion of the general long-range spreading,
Eq.\ (\ref{Inf-Spread}). Hence, we expect the same difficulties concerning the positivity of the propagator for momenta $q>\mu$ and $\sigma>2$ if $v>2\alpha$.

The RG functions for the new variables are easily derived from
Eq.\ (\ref{Def-RGFu-overbar}) by using Eqs.\ (\ref{Gamma-Ident}),
\begin{align}
\beta_{u}  &  =\left.  \mu\frac{\partial u}{\partial\mu}\right\vert
_{\mathrm{bare}}=\left(  -\varepsilon+2v+2\gamma_{\lambda}+\gamma-\gamma
_{u}\right)  u\,,
\nonumber\\
\beta_{v} &=\left.  \mu\frac{\partial v}{\partial\mu
}\right\vert _{\mathrm{bare}}=\left(  -2\alpha+v+\gamma_{\lambda}\right)
v\,,
\nonumber\\
\zeta &  =\left.  \mu\frac{\partial\ln\lambda}{\partial\mu}\right\vert
_{\mathrm{bare}}=\gamma-\gamma_{\lambda}-v\, ,
\nonumber\\
\kappa&=\left.  \mu
\frac{\partial\ln\tau}{\partial\mu}\right\vert _{\mathrm{bare}}=\gamma
_{\lambda}-\gamma_{\tau}+v\,. \label{RGFu-hybr}%
\end{align}
In comparison with the short-range RG functions, Eq.~(\ref{Def-RGFu}), the
new Gell-Mann--Low function $\beta_{v}$ as well as the functions $\beta_{u}$,
$\zeta$, and $\kappa$ have only an additive contribution of the variable $v$.
Note that we have to consider both $u$ and $v$ as being parameters of
order $\varepsilon\sim\alpha$. In terms of the new variables, the RGE is still of the form given in Eq.~(\ref{RGE-SR}), however, here
\begin{equation}
\mathcal{D}=\mu\frac{\partial}{\partial\mu}+\zeta\lambda\frac{\partial
}{\partial\lambda}+\kappa\tau\frac{\partial}{\partial\tau}+\beta_{u}%
\frac{\partial}{\partial u}+\beta_{v}\frac{\partial}{\partial v}\, ,
\label{RG-Op-hybr}%
\end{equation}
has to  be inserted as the RG differential operator.

\subsection{Asymptotic scaling regions in $(\varepsilon,\alpha)$-expansion}
\label{sec:hybridModeScalingRegions}

In this subsection we analyze the different scaling regions in a $(d,\sigma
)$-diagram using the results on the hybrid model derived in Sec.~\ref{sec:hybridModelRen}.
The asymptotic scaling of response and correlation functions
is governed by the various fixed points of the renormalization group. To find stable fixed points of
the RGE and the corresponding scaling behavior, we have to find solutions of
the equations $\beta_{u}=\beta_{v}=0$ with the Gell-Mann--Low functions $\beta_{u}$ and $\beta_{v}$ as
given in Eqs.~(\ref{RGFu-hybr}). The different fixed points can be classified
by setting to zero the different factors of these functions.

There is the trivial fixed point $u_{\ast}=v_{\ast}=0$. To find its stability
conditions we determine the eigenvalues of its stability matrix%
\begin{equation}
\underline{\underline{\beta_{\ast}}}=%
\begin{pmatrix}
\left.  \partial\beta_{u}/\partial u\right\vert _{\ast} & \left.
\partial\beta_{u}/\partial v\right\vert _{\ast}\\
\left.  \partial\beta_{v}/\partial u\right\vert _{\ast} & \left.
\partial\beta_{v}/\partial v\right\vert _{\ast}%
\end{pmatrix}
=%
\begin{pmatrix}
-\varepsilon & 0\\
0 & -2\alpha
\end{pmatrix}
\,. \label{Stab-triv}%
\end{equation}
The eigenvalues of this matrix, $\omega_{1}=-\varepsilon$ and
$\omega_{2}=-2\alpha$, are positive for $d>4$ and $\sigma>2$, i.e., we retrieve the
mean-field region of short-range DP.

Another fixed point, the trivial long-range fixed point, is given by $u_{\ast}=0$, $v_{\ast}=2\alpha$. Its
stability matrix reads
\begin{equation}
\underline{\underline{\beta_{\ast}}}=%
\begin{pmatrix}
4\alpha-\varepsilon & 0\\
2\alpha\gamma_{\lambda\ast}^{\prime} & 2\alpha
\end{pmatrix}
\,, \label{Stab-triv-LR}%
\end{equation}
where the stroke at $\gamma_{\lambda}^{\prime}$ denotes the derivative with respect to $u$.
The eigenvalues $\omega_{1}=4\alpha-\varepsilon$ and $\omega_{2}=2\alpha$ are positive for  $d>4(1-\alpha)=2\sigma$, $\sigma<2$ which marks the region of stability of the trivial long-range fixed point.

Next we come to the fixed point $v_{\ast}=0$, $u_{\ast}>0$, as the solution of
$2\gamma_{\lambda\ast}+\gamma_{\ast}-\gamma_{u\ast}=\varepsilon$. Of course,
this is the fixed point of the normal short-range DP with $u_{\ast}=u_{\ast
}^{DP}$. The stability matrix is%
\begin{equation}
\underline{\underline{\beta_{\ast}}}=%
\begin{pmatrix}
(2\gamma_{\lambda\ast}^{\prime}+\gamma_{\ast}^{\prime}-\gamma_{u\ast}^{\prime
})u_{\ast} & 2u_{\ast}\\
0 & \gamma_{\lambda\ast}-2\alpha
\end{pmatrix}
\,. \label{Stab-DP-SR}%
\end{equation}
The first eigenvalue of this matrix, $\omega_{1}$=$(2\gamma_{\lambda\ast
}^{\prime}+\gamma_{\ast}^{\prime}-\gamma_{u\ast}^{\prime})u_{\ast}%
=\varepsilon+O(\varepsilon^{2})$ shows the stability range of non-trivial
short-range DP: $d<4$. Using $\gamma_{\lambda_{\ast}}=\gamma_{\lambda
}(u_{\ast}^{DP})=\eta^{SR}+2-z^{SR}$, we find the stability condition against
long-range spreading:
\begin{equation}
\sigma>z^{SR}-\eta^{SR}\,. \label{Stab-DP-SR-Bdg}%
\end{equation}

Now, we come to the interesting LR-region where $u_{\ast}>0$ and $v_{\ast}%
\neq0$. In this domain, the stable fixed points of (\ref{RGFu-hybr}) are
solutions of the fixed point equations
\begin{equation}
v_{\ast}=2\alpha-\gamma_{\lambda\ast}>0\ ,\quad\bar{\varepsilon}%
=\varepsilon-4\alpha=\gamma_{\ast}-\gamma_{u\ast}\, . \label{FP-Gl-LR}%
\end{equation}
Using the $\gamma$-functions which follow from Eq.~(\ref{Z_SR}), and which have been utilized in
Eq.~(\ref{RGE-Fu-SR}), we find the fixed point
\begin{align}
u_{\ast}^{LR}  &  =\frac{4\bar{\varepsilon}}{7}\left[  1+\left(  50+9\ln
\frac{4}{3}\right)  \frac{\bar{\varepsilon}}{98}+O\left(  \bar{\varepsilon
}^{2},\alpha\bar{\varepsilon}\right)  \right]  ,
\nonumber\\
v_{\ast}^{LR}  &  =2\alpha+\frac{\bar{\varepsilon}}{14}\left[  1-\left(
17-526\ln\frac{4}{3}\right)  \frac{\bar{\varepsilon}}{392}+O\left(
\bar{\varepsilon}^{2},\alpha\bar{\varepsilon}\right)  \right] .
\label{uw-LR}%
\end{align}
The critical exponents in the LR-region are found from Eq.\ (\ref{RGFu-hybr})
as $\eta_{LR}=\gamma(u_{\ast}^{LR})$ and $1/\nu_{LR}=2-\kappa(u_{\ast}^{LR})$
with the expansions%
\begin{align}
\eta_{LR}  &  =-\frac{\bar{\varepsilon}}{7}\left[  1+c_{\eta}(\alpha
)\bar{\varepsilon}+O\left(  \bar{\varepsilon}^{2}\right)  \right]  \,,\
\nonumber\\
1/\nu_{LR}  &  =\sigma-\frac{2\bar{\varepsilon}}{7}\left[  1+c_{\nu}%
(\alpha)\bar{\varepsilon}+O\left(  \bar{\varepsilon}^{2}\right)  \right]
\,, \label{Exp-LR-exp-hyb}%
\end{align}
where
\begin{align}
c_{\eta}(\alpha)&=\left(  \frac{4}{49}+\frac{36}{49}\ln\frac{4}{3}\right)
+O\left(  \alpha\right)  \,,
\nonumber\\
c_{\nu}(\alpha) &=\left(  \frac{15}{98}+\frac{9}{98}\ln\frac{4}%
{3}\right)  +O\left(  \alpha\right)  \, .%
\end{align}
Note that we have already encountered the first order contributions
in Eq.~(\ref{Exp-LR-exp}) above. Of course, $z_{LR}$ follows from the exact relation
(\ref{LR-exRel}). The stability matrix for this fixed point reads
\begin{equation}
\underline{\underline{\beta_{\ast}}}=%
\begin{pmatrix}
Au_{\ast} & 2u_{\ast}\\
\gamma_{\lambda\ast}^{\prime}v_{\ast} & v_{\ast}%
\end{pmatrix}
\,, \label{Stab-LR}%
\end{equation}
with $A=(2\gamma_{\lambda\ast}^{\prime}+\gamma_{\ast}^{\prime}-\gamma_{u\ast
}^{\prime})=3/2+O(\bar{\varepsilon})>0$. The two eigenvalues are given by%
\begin{equation}
\omega_{\pm}=\left(  \frac{Au_{\ast}+v_{\ast}}{2}\right)  \pm\sqrt{\left(
\frac{Au_{\ast}+v_{\ast}}{2}\right)  ^{2}-\left(  A-2\gamma_{\lambda\ast
}^{\prime}\right)  u_{\ast}v_{\ast}}\,. \label{Eigv-LR}%
\end{equation}
They are positive as long as $\left(  A-2\gamma_{\lambda\ast}^{\prime}\right)
u_{\ast}v_{\ast}=(\gamma_{\ast}^{\prime}-\gamma_{u\ast}^{\prime})u_{\ast
}v_{\ast}=(49/16+O(\bar{\varepsilon}))\bar{\varepsilon}v_{\ast}>0$. This
condition leads to $\bar{\varepsilon}>0$ and $v_{\ast}=2\alpha-\gamma
_{\lambda\ast}>0$. The long-range fixed point looses its stability if the line
$2\alpha=\gamma_{\lambda\ast}$ is reached. At that point, all critical
exponents change over continuously to the usual short-range DP-exponents as
can be easily seen from Eq.\ (\ref{RGFu-hybr}). Hence, stability boundary of
the long-range L\'{e}vy-flight exponent $\sigma=2(1-\alpha)$ is given by%
\begin{align}
\sigma&=\sigma_{c}=z_{SR}-\eta_{SR}
\nonumber\\
&=2-\frac{\varepsilon}{12}\left[  1+\left(
-\frac{17}{288}+\frac{263}{144}\ln\frac{4}{3}\right)  \varepsilon+O\left(
\varepsilon^{2},\alpha\varepsilon\right)  \right]  \,, \label{StabBound}%
\end{align}
which is less than two. This fact is astonishing because for $\sigma$ lower
then $2$ but greater then $\sigma_{c}$, the long-range part $q^{\sigma}$ is
naively irrelevant in comparison to the normal diffusional part $q^{2}$.
However, in an interacting theory it is not the free propagator but rather the
response function $\chi(\omega,\mathbf{q};\tau)=\Gamma_{1,1}(\omega
,\mathbf{q};\tau)^{-1}=q^{\eta-z}\,f\left(  \omega/q^{z},\tau/q^{1/\nu
}\right)  $ which is the deciding quantity. Hence, one has to compare
$q^{\sigma}$ with $q^{z_{SR}-\eta_{SR}}$ to find out which is leading for
$q\to0$. The other stability boundary is approached if $u_{\ast}$ goes to zero, and is
given by $\bar{\varepsilon}=\varepsilon-4\alpha=0$. This boundary coincides
with the value found above: $d=2\sigma$. At this line the exponents cross
over to their long-range mean-field values.

To summarize our findings regarding the scaling regions: all
boundaries between the four scaling regions, namely short-ranged DP, long-ranged DP,
as well their two mean-field counterparts, are generally given by the four lines
in a $(d,\sigma)$-diagram where one or two of the fixed point values $u_{\ast
}$ and $v_{\ast}$ vanish. There is no room for other stability lines as some authors
argued~\cite{Antonov}.

\subsection{Landau's ghost}
\label{sec:landauGhost}
As we have remarked at several points, the propagator becomes problematic for higher momenta. Consider the $q$-dependent part of the inverse renormalized propagator of the hybrid theory,
\begin{equation}
\label{prop}
G\left(\mathbf{q},0,0\right)^{-1}=\lambda q^{2}\left[1+\frac{v}{2\alpha}\left((q/\mu)^{-2\alpha}
-1\right) \right] ,
\end{equation}
where $v\geq 0$ to ensure stability (positivity) for $q \to 0$ for both signs of $\alpha$. The part proportional to $v$ changes sign and leads to a loss of stability for negative $\alpha$ at a momentum $q_g$ given by
\begin{equation}
\ln(q_g/\mu)=\frac{1}{2|\alpha|}\ln(1+2|\alpha|/v) \sim 1/v
\end{equation}
for small $\alpha$. This ghost reminds of Landau's ghost in quantum electrodynamics. The momentum of the ghost goes exponentially to infinity if $v \to 0$. This ghost even arises for positive $\alpha$ if $v>2\alpha$. The fixed point of the hybrid theory does always belong to this region! The correct interpretation is the following: our asymptotic theory is just an effective theory in the sense that it can only be used in the low momentum limit in a perturbation expansion with $v$ as an expansion parameter. Therefore, the second part of Eq.~(\ref{prop}) must be considered as a perturbation. 

Let us demonstrate this in some detail for the inverse response function $\Gamma_{1,1}$ as an example. To this end, we work to 1-loop order using $u$, $v$, $\varepsilon$, and $\alpha$ as first order quantities.   The zeroth order is
\begin{equation}
\Gamma_{1,1}^{(0)}\left(  \mathbf{q},\omega=0,\tau=0\right)  =\lambda q^{2}\,.
\end{equation}
Using renormalized perturbation theory, adding both first order terms, and neglecting higher order terms, we obtain
\begin{align}
\Gamma_{1,1}&\left(  \mathbf{q},0,0\right) \nonumber\\ 
& =\lambda q^{2}\left[ 1+ \frac{v}{2\alpha}\left(  \left(  \frac{q}{\mu
}\right)  ^{-2\alpha}-1\right)  +\frac{u}{8}\ln\left(  \frac{q}{2\mu}\right)
\right]  +\ldots\nonumber\\
&  =\lambda q^{2}\left[  1-v\ln\left(  \frac{q}{\mu}\right)  +\frac{u}{8}%
\ln\left(  \frac{q}{2\mu}\right)  \right]  +\ldots\nonumber\\
&  =\left(  1-\frac{u}{8}\ln2\right)  \lambda q^{2}\left[  1+\left(  \frac
{u}{8}-v\right)  \ln\left(  \frac{q}{\mu}\right)  \right]  +\ldots
\end{align}
Now we use the fixed point result $v_{\ast}=2\alpha-\gamma_{\lambda\ast}$ with
$\gamma_{\lambda}=-u/8$,  to get%
\begin{align}
\Gamma_{1,1}&\left(  \mathbf{q},\omega=0,\tau=0\right) \nonumber\\ 
&=\left(  1-\frac{u_{\ast}}{8}\ln2\right)  \lambda q^{2}\left[  1-2\alpha\ln\left(  
\frac{q}{\mu}\right)  \right]  +\ldots \nonumber\\
&\sim\lambda q^{2\left(  1-\alpha\right)  }\, .
\end{align}
As expected, the RG proves to be the systematic tool to resum all logarithms to yield the correct critical exponent.

This procedure holds for all $\alpha=O\left(  \varepsilon\right)$
irrespective of the sign as long as the fixed point with $v_{\ast}>0$ is
stable. Otherwise $v_{\ast}=0$, $u_{\ast}=2\varepsilon/3$, and one gets
$\Gamma_{1,1}\sim\lambda q^{2-\eta_{SR}}$ with $\eta_{SR}=-\varepsilon/12$, i.e.,
the known behavior for the short-range case. We once more point out that the second term (the
L\'{e}vy-flight contribution) has to be handled as a perturbation to the
desired order and not as a part of the unperturbed propagator. Also, one has to
interpret the special case $\alpha=0$ as a relevant logarithmic perturbation
$\sim vq^{2}\ln\left(  \mu/q\right)  $ of $q^{2}$. Only if $v$ is strictly
zero the short range case is recovered. Thus, one cannot expect a continuous
behavior at $\alpha=0$ comparing the short-range versus the L\'{e}vy-flight
directed percolation.

\section{Critical behavior of dynamic observables in L\'{e}vy-flight directed percolation}
\label{sec:criticalBehavior}

In this section we will harvest some of our previous results to calculate scaling forms and logarithmic correction for those dynamic quantities in long-ranged DP that are most suitable from the vantage point of numerical simulations~\cite{Hin07}. Two key observables with respect to simulations are the density of
infected individuals $\rho(t)=\langle s(\mathbf{r},t)\rangle_{\rho_{0}}$ for
$t>0$ if the initial state at time $t=0$ is prepared with a homogeneous
initial density $\rho_{0}$, and the response function $\chi(\mathbf{r}%
,t)=\langle s(\mathbf{r},t)\tilde{s}(\mathbf{0},0)\rangle$ that yields the
density of infected individuals after the epidemic is initialized by a
pointlike source at $t=0$ and $\mathbf{r}=0$.

\subsection{Scaling properties}

The scaling properties of the density of infected individuals and the response function follow
from the RGE (\ref{RGE-SR}) taken at the long-range fixed point of
Eq.~(\ref{RGFu-hybr}) and by identifying $\chi(\mathbf{r},t)$ and $\rho(t)$ with the
Green functions $G_{1,1}(\mathbf{r},t;\tau)$ and $G_{1,0}(\mathbf{r},t;\tau,\rho_{0})$, respectively. The initial density $\rho_{0}$ is introduced
into the the response functional via a (bare) source $\mathring{h}%
(\mathbf{r},t)=\mathring{\rho}_{0}\delta(t)$ with renormalization
$\mathring{\rho}_{0}=Z^{-1/2}\rho_{0}$ \cite{JaTa05}, which leads to an
additional derivative term $\frac{1}{2}\gamma\rho_{0}\partial/\partial\rho
_{0}$ in the RGE. We obtain the scaling forms%
\begin{align}
\rho(t)  &  =t^{-\delta}S_{\rho}(\tau t^{1/z\nu},\rho_{0}t^{\delta+\theta
})\,,\nonumber\\
\chi(\mathbf{r},t)  &  =t^{-2\delta}S_{\chi}(\mathbf{r}/t^{1/z};\tau
t^{1/z\nu})\,, \label{Scal-Obs}%
\end{align}
where the $S_{...}$ are appropriate scaling functions. We drop in
this section all the subscripts at the critical exponents because we are
interested in the long-range case only. Hence, $\eta=\eta_{LR}$ and
$1/\nu=1/\nu_{LR}$, see Eqs.\ (\ref{Exp-LR-exp-hyb}), and $z=\sigma+\eta$,
$\delta=(d+\eta)/2z$, $\theta=-\eta/z$. The expansions of the latter two are%
\begin{align}
\delta &  =1-\frac{3\bar{\varepsilon}}{7\sigma}\left[  1+c_{\delta}%
(\alpha)\bar{\varepsilon}+O\left(  \bar{\varepsilon}^{2}\right)  \right]\,,
\nonumber\\
\theta &  =\frac{\bar{\varepsilon}}{7\sigma}\left[  1+c_{\theta}(\alpha
)\bar{\varepsilon}+O\left(  \bar{\varepsilon}^{2}\right)  \right]  \,,
 \label{Exp-LR-exp-hyb-2}%
\end{align}
where
\begin{align}
c_{\delta}(\alpha)&=\left(  \frac{17}{294}-\frac{6}{49}\ln\frac{4}%
{3}\right)  +O\left(  \alpha\right)  \,,
\nonumber\\
c_{\theta}(\alpha)&=\left(  \frac{23}{196}+\frac{36}{49}\ln\frac{4}{3}\right)
+O\left(  \alpha\right)\,.
\end{align}
At the critical point $\tau=0$, Eqs.~(\ref{Scal-Obs}) show that the mean
square radius of spreading from the origin scales as $R^{2}(t)\sim t^{2/z}$,
and the average number of infected individuals $N(t)=\int d^{d}r\chi
(\mathbf{r},t)\sim t^{\theta}$. Starting with a homogeneous finite value
$\rho_{0}$, the critical density first increases in a universal time regime
with the same exponent as $\rho(t)\sim\rho_{0}t^{\theta}$. Then, after some
crossover time, it decreases as $\rho(t)\sim t^{-\delta}$. If one starts with
a full lattice of infected sites corresponding to an infinite initial value
$\rho_{0}$, only the last scaling behavior is seen. Because of asymptotic
time-reflection invariance of DP (duality symmetry) this behavior
characterizes also the survival probability \cite{Ja05}, $P(t)\sim t^{-\delta}$.

\begin{figure}
\centerline{\includegraphics[width=8.4cm]{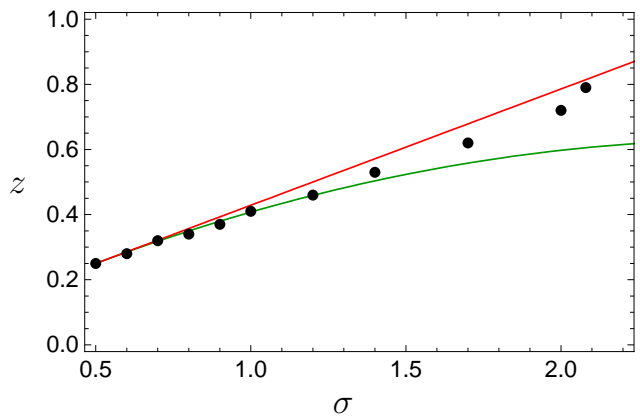}}
\caption{(Color online) The exponent $z$ as a function of $\sigma$ for $d=1$. The data points stem from simulations by data by Hinrichsen~\cite{Hin01}. The red (upper) and the green (lower) curve correspond to our 1-loop and 2-loop results, respectively.}
\label{fig:z}
\end{figure}

\begin{figure}
\centerline{\includegraphics[width=8.4cm]{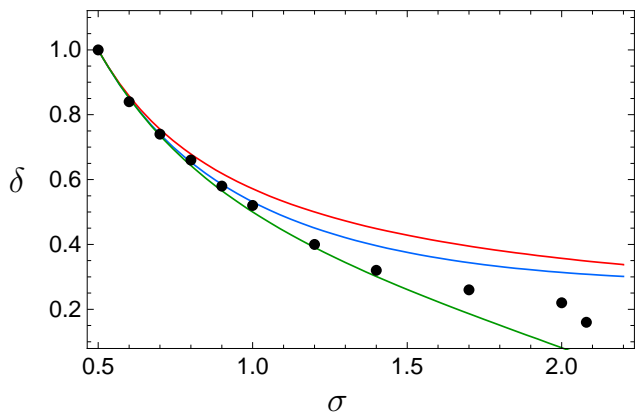}}
\caption{(Color online) The exponent $\delta$ as a function of $\sigma$ for $d=1$. For an explanation of the curves, see the main text.}
\label{fig:delta}
\end{figure}

\begin{figure}
\centerline{\includegraphics[width=8.4cm]{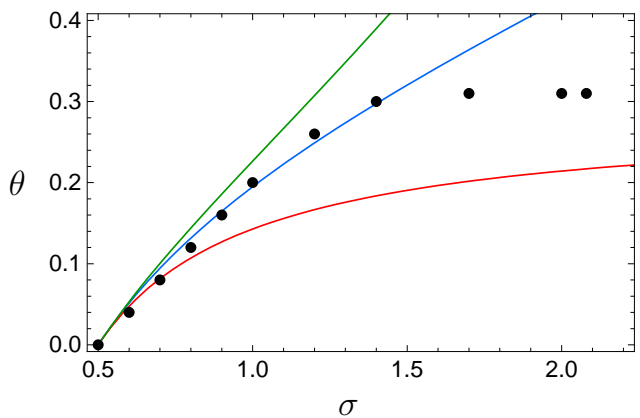}}
\caption{(Color online) The exponent $\theta$ as a function of $\sigma$ for $d=1$. The color-coding of the curves is the same as in Fig.~\ref{fig:delta}.}
\label{fig:theta}
\end{figure}

To first order all exponents are identical to the LR-exponents
derived from Eqs.~(\ref{Exp-LR-exp}), of course. The full $\alpha$-content of the
functions $c_{...}(\alpha)$ in Eqs.~(\ref{Exp-LR-exp-hyb}) and (\ref{Exp-LR-exp-hyb-2}) must be calculated from the long-range model,
Eq.~(\ref{J-LR}). Hence, one has to be careful when applying the expansions
to $O\left(  \bar{\varepsilon}^{2}\right)$ for $\alpha=0$, e.g., in $d=1$ where $\alpha
=(3-\bar{\varepsilon})/4$. In Figs.~\ref{fig:z} to \ref{fig:theta}, we compare our results for the exponents $z$, $\delta$ and $\theta$ to numerical results for $d=1$ by Hinrichsen~\cite{Hin01}. For this comparison, we use the first order expansions in $\bar{\varepsilon}$ (red curves), which are exact in
$\alpha$, and the second order expansions in $\bar{\varepsilon}$ where we neglect the $\alpha$ dependent parts of the $c_{...}(\alpha)$ (green and blue curves). The green curves show our second order result for $z$ and results obtained for $\delta$ and $\theta$ by using without further expansion the scaling relations relating $\delta$ and $\theta$ to $z$ and $\eta$. The blue curves stem from using these scaling relations and then properly expanding $\delta$ and $\theta$ to second order in $\bar{\varepsilon}$. For $\sigma$ in the range from $1/2$ to roughly $1$, the numerical data and the analytic results agree remarkably well. For larger $\sigma$ sigma, the agreement suffers, but is well within the expectations for the methods used here.

\subsection{Logarithmic corrections}

Above the boundary between the genuine and the trivial long-range regions
($d>2\sigma$, $\sigma<2$), the coupling constant $g$ tends to zero under the RG. However, $g$ represents a dangerously
irrelevant variable here, since it scales various observables, and setting
$g=0$ rigorously leads either to zero or infinity for relevant quantities. Due to its
twofold nature as both a relevant scaling variable and an irrelevant
loop-expansion generating parameter, $g$ has to be treated very carefully. To set the stage for such a treatment, let us briefly review a few fundamentals of dynamic field theory.
In broad terms, one attempts to determine the cumulant generating functional defined by the functional integral%
\begin{align}
\mathcal{W}_{\mathrm{LR}}\left[  H,\tilde{H}\right]  &=\ln\int\mathcal{D}%
\left[  \tilde{s},s\right]  \exp\Big[  -\mathcal{J}_{\mathrm{LR}}\left[
\tilde{s},s\right]  +\left(  H,s\right)
\nonumber\\
&+(  \tilde{H},\tilde{s})
\Big]  \,.\label{KumGen}%
\end{align}
Functional derivatives with respect to the sources $H$ and $\tilde{H}$ define
the Green's functions. The generating functional for the vertex functions $\Gamma_{\mathrm{LR}}\left[  \tilde{s},s\right] $, the dynamic free energy, is related to the cumulant generating functional via the Legendre transformation%
\begin{equation}
\Gamma_{\mathrm{LR}}\left[  \tilde{s},s\right]  +\mathcal{W}_{\mathrm{LR}%
}\left[  H,\tilde{H}\right]  =\left(  H,s\right)  +\left(  \tilde{H},\tilde
{s}\right) ,  \label{Leg-Tr}%
\end{equation}
with $s=\delta\mathcal{W}_{\mathrm{LR}}/\delta H$ and $\tilde{s}%
=\delta\mathcal{W}_{\mathrm{LR}}/\delta\tilde{H}$, and vice versa.
In terms of $\Gamma_{\mathrm{LR}}$, the twofold nature of $g$ is lucidly exposed by writing
\begin{equation}
\Gamma_{\mathrm{LR}}[\tilde{s},s;\tau,g]=g^{-2}\Phi_{\mathrm{LR}}[g\tilde
{s},gs;\tau,u]\,. \label{lucid-Gamma}%
\end{equation}
The expansion of the functional $\Phi_{\mathrm{LR}}[g\tilde{s},gs;\tau,u]$
into a series with respect to $u$ yields the loop expansion. The zeroth term
$g^{-2}\Phi_{\mathrm{LR}}[g\tilde{s},gs;\tau,0]$ is just the response
functional $\mathcal{J}_{\mathrm{LR}}$, Eq.~(\ref{J-LR}), itself. The scaling
form of the generating functional for the cumulants that corresponds to
Eq.~(\ref{lucid-Gamma}) reads%
\begin{equation}
\mathcal{W}_{\mathrm{LR}}[H,\tilde{H};\tau,g]=g^{-2}\Omega_{\mathrm{LR}%
}[gH,g\tilde{H};\tau,u]\,. \label{lucid-W}%
\end{equation}
To leading order in the logarithmic corrections, we may neglect the dependence
of $\Omega$ and $\Phi$ on $u$. Functional derivation lead the to the Green's
functions%
\begin{equation}
G_{N,\tilde{N}}(\{\mathbf{r},t\},\tau;u)\simeq u^{-1+(N+\tilde{N}%
)/2}T_{N,\tilde{N}}\left(  \{\mathbf{r},t\},\tau,u^{1/2}\rho_{0}\right) ,
\end{equation}
where $T_{N,\tilde{N}}$ are the contributions of loopless trees consisting of
$N+\tilde{N}-1$ propagators and $N+\tilde{N}-2$ vertices.

The characteristic equations that follow from the RG-functions, Eqs.~(\ref{RGFu-LR-exp}), with $\bar{\mu}(\ell)=\ell\mu$ and $d=2\sigma$, are to lowest order given by
\begin{align}
\ell\frac{d\bar{u}(\ell)}{d\ell}  &  =\bar{\beta}(\bar{u}(\ell))=\frac{7}%
{4}\bar{u}(\ell)^{2}\,,
\nonumber\\
\frac{d\ln X(\ell)}{d\ln\ell}&=\bar{\gamma}(\bar
{u}(\ell))=-\frac{\bar{u}(\ell)}{4}\,,
\nonumber\\
\frac{d\ln X_{\lambda}(\ell)}{d\ln\ell}  &  =\bar{\zeta}(\bar{u}(\ell
))=-\frac{\bar{u}(\ell)}{4}\,,
\nonumber\\
\frac{d\ln X_{\tau}(\ell)}{d\ln\ell}%
&=\bar{\kappa}(\bar{u}(\ell))=\frac{\bar{u}(\ell)}{2}\,, \label{Char-Eq}%
\end{align}
Solving these equations, we obtain asymptotically for $\ell\ll1$,
\begin{align}
\bar{u}(\ell)  &  \sim\left\vert \ln\ell\right\vert ^{-1}\,,\quad X(\ell
)\sim\left\vert \ln\ell\right\vert ^{1/7}\,,\nonumber\\
X_{\lambda}(\ell)  &  \sim\left\vert \ln\ell\right\vert ^{1/7}\,,\quad
X_{\tau}(\ell)\sim\left\vert \ln\ell\right\vert ^{-2/7}\,. \label{As-Char}%
\end{align}
Hence, we get
\begin{align}
&G_{N,\tilde{N}}(\{\mathbf{r},t\},\tau;u)
\nonumber\\
&  \simeq\bar{u}(\ell)^{-1}\left[
\bar{u}(\ell)\ell^{2\sigma}X(\ell)\right]  ^{(N+\tilde{N})/2}\nonumber\\
&  \times T_{N,\tilde{N}}\left(  \{\ell\mathbf{r},\ell^{\sigma}X_{\lambda
}(\ell)t\},\ell^{-\sigma}X_{\tau}(\ell)\tau,\bar{u}(\ell)^{1/2}\ell^{-\sigma
}X(\ell)^{1/2}\rho_{0}\right) \nonumber\\
&  \sim\left\vert \ln\ell\right\vert \,\left[  \ell^{\sigma}\left\vert \ln
\ell\right\vert ^{-3/7}\right]  ^{N+\tilde{N}}\nonumber\\
&  \times T_{N,\tilde{N}}\left(  \{\ell\mathbf{r},\ell^{\sigma}\left\vert
\ln\ell\right\vert ^{1/7}t\},\ell^{-\sigma}\left\vert \ln\ell\right\vert
^{-2/7}\tau,\ell^{-\sigma}\left\vert \ln\ell\right\vert ^{-3/7}\rho
_{0}\right)  \label{As-Green}%
\end{align}
as solutions of the entire RGE . Choosing either $\ell^{\sigma}\left\vert \ln\ell\right\vert ^{1/7}t\sim1$ or
$\ell^{-\sigma}\left\vert \ln\ell\right\vert ^{-2/7}\tau\sim1$, we deduce
that at the critical point
\begin{align}
\rho(t)  &  \sim\rho_{0}\,\left(  \ln t\right)  ^{1/7}
\end{align}
in the initial time region,
\begin{align}
\rho(t)  &  \sim P(t)\sim t^{-1}\,\left(  \ln t\right)  ^{3/7}
\end{align}
in the late time region, and
\begin{align}
R^{2}(t)  &  \sim\left[  t\,\left(  \ln t\right)  ^{1/7}\right]  ^{2/\sigma
}\,. \label{LogKorrekt1}%
\end{align}

\section{Concluding remarks}
\label{sec:concludingRemarks}
In summary, we studied DP with L\'{e}vy-flight spreading by using the powerful methods of renormalized field theory. Our work confirms the previously known RG fixed point structure including their stability regions and the fact that the critical exponents change continuously in the crossover between short-range DP and L\'{e}vy DP. We calculated the critical exponents for L\'{e}vy DP, which have hitherto been known to first order, to second order in an expansion in $\varepsilon$ and $\alpha$. These results agree well with the existing numerical simulations for $d=1$. In addition, we calculated the leading logarithmic corrections for several dynamical observables that are typically measured in simulations.

We hope that our work stimulates further interest in long-range DP. It would be interesting to see further simulation results, e.g., for the critical exponents for $d>1$ and for logarithmic corrections. Also, it would be interesting to have analytical and numerical results for other universal quantities such as scaling functions and amplitudes. In a forthcoming paper we will apply the same methods to the long-range GEP, that is to dynamic isotropic percolation.

\begin{acknowledgments}
  This work was supported in part (O.S.) by the National Science Foundation under
  grant No.\ DMR 0804900.
\end{acknowledgments}

\end{document}